\documentclass[a4paper,twocolumn,showpacs,superscriptaddress,floatfix,prl,nofootinbib]{revtex4-1}
\usepackage{graphicx}
\usepackage{epsf}
\usepackage{epsfig}
\usepackage{amssymb,amsmath}
\usepackage[usenames]{color}
\usepackage{amssymb}
\usepackage{times}
\usepackage{hyperref}
\newcommand{\cf}{\textit{cf.}~}
\newcommand{\ie}{\textit{i.e.}~}
\newcommand{\eg}{\textit{e.g.}~}

\begin{document}

\title{Will black hole-neutron star binary inspirals tell us about
   the neutron star equation of state?}

\author{Francesco Pannarale}
\affiliation{
Max-Planck-Institut f{\"u}r Gravitationsphysik, Albert Einstein
Institut, Potsdam, Germany 
}

\author{Luciano~Rezzolla}
\affiliation{
Max-Planck-Institut f{\"u}r Gravitationsphysik, Albert Einstein
Institut, Potsdam, Germany
}
\affiliation{
Department of Physics and Astronomy,
Louisiana State University,
Baton~Rouge, LA, USA
}

\author{Frank~Ohme}
\affiliation{
Max-Planck-Institut f{\"u}r Gravitationsphysik, Albert Einstein
Institut, Potsdam, Germany
}

\author{Jocelyn~S.~Read}
\affiliation{
Department of Physics and Astronomy, The University of Mississippi, University, MS 38677, USA
}

\begin{abstract}
  The strong tidal forces that arise during the last stages of the
  life of a black hole-neutron star binary may severely distort, and
  possibly disrupt, the star. Both phenomena will imprint signatures
  about the stellar structure in the emitted gravitational
  radiation. The information from the disruption, however, is confined
  to very high frequencies, where detectors are not very sensitive. We
  thus assess whether the lack of tidal distortion corrections in
  data-analysis pipelines will affect the detection of the
  \emph{inspiral} part of the signal and whether these may yield
  information on the equation of state of matter at nuclear
  densities. Using recent post-Newtonian expressions and realistic
  equations of state to model these scenarios, we find that
  point-particle templates are sufficient for the detection of black
  hole-neutron star inspiralling binaries, with a loss of signals
  below $1\%$ for both second and third-generation detectors. Such
  detections may be able to constrain particularly stiff equations of
  state, but will be unable to reveal the presence of a neutron star
  with a soft equation of state.
\end{abstract}

\maketitle

\noindent\emph{Introduction.~}Coalescing binaries of stellar-mass
compact objects, \ie black holes (BHs) and neutron stars (NSs), are a
primary target for gravitational wave (GW) searches performed with
kilometer-size laser-interferometric detectors, such as LIGO and
Virgo, and one of the most promising for a first detection. Binaries
containing at least one NS are particularly interesting because their
GW signal contains signatures of the physical conditions of matter at
nuclear densities (see, \eg~\cite{Baiotti08, Duez09, Kyutoku2010,
  Giacomazzo:2009mp, Damour:2009, Hinderer09, Damour:2009wj,
  Rezzolla:2010, Baiotti:2010}) and may thus eventually reveal the
equation of state (EOS) of NSs, which is currently highly
uncertain~\cite{Lattimer07}.

In the case of BH-NS binaries, in particular, the most relevant of
such signatures was traditionally thought to be the GW frequency above
which no hydrostatic equilibrium is possible, $f_{\rm tide}$. Because
this frequency was identified with the complete tidal disruption of
the NS, the GW amplitude was consequently assumed to decay rapidly for
larger frequencies~\cite{Vallisneri00} and information on the EOS was
thought to be inferrable from $f_{\rm tide}$ for binaries in which the
NS were to be disrupted before the system reaches the Innermost Stable
Circular Orbit (ISCO). This picture was considerably modified by the
recent numerical-relativity simulations presented
in~\cite{ShibataTaniguchi2008} which, instead, highlighted that
$f_{\rm tide}$ effectively marks only the onset of
\emph{mass-shedding}. The complete tidal disruption, on the other
hand, is achieved only later and at higher frequencies, so that the GW
spectrum decays exponentially at a cutoff frequency \hbox{$f_{\rm
    cutoff} \simeq (1.2-1.5) f_{\rm tide}$}. Stated differently, in
contrast to the expectations of~\cite{Vallisneri00}, the GW spectrum
shows no distinctive feature at the mass-shedding frequency $f_{\rm
  tide}$; conversely, information on the NS EOS is in principle
accessible through $f_{\rm cutoff}$.  The two frequencies do not
coincide because the tidal disruption is not instantaneous and the GW
frequency changes rapidly in time, so that the transition from a
chirping signal to an exponentially decaying one is pushed to higher
frequencies. The difference between $f_{\rm tide}$ and $f_{\rm
  cutoff}$ increases as $f_{\rm tide}$ approaches the GW frequency at
the ISCO, $f_{\text{ISCO}}$~\cite{ShibataTaniguchi2008}.

If recent numerical relativity simulations have pointed out the
importance of the cutoff frequency as a very significant marker of the
NS properties, they have also highlighted that the accurate
determination of such frequency may only be made through numerical
simulations, the computational costs of which are still prohibitive
for a complete scan of the space of parameters. Moreover, it will be
the inspiral portion of the GW signal that will be fully contained in
the most sensitive part of the sensitivity curve of ground-based
detectors.

In light of all this, two questions follow naturally. (1) Will the use
of point-particle template banks for \textit{inspiral} searches cause
a loss in the number of detected signals because of the small but
secular tidal effects that develop? (2) Will such effects be large
enough to provide us with information about the NS EOS? We address
here both questions by computing the GW phase evolution of BH-NS
binaries when tidal effects are taken into account and by contrasting
it with the one expected when both the BH and NS are treated as
point-particles. We find that the tidal corrections to the phase
evolution depend sensitively on the NS EOS, but also that they are
generally small. Hence, present data-analysis pipelines, which do not
include tidal corrections, are sufficient for a successful detection
of BH-NS inspirals, but only a particularly stiff EOS would mark the
GW signal sufficiently for us to determine it.

%FFFFFFFFFFFFFFFFFFFFFFFFFFFFFFFFFFFFFFFFFFFFFFFFFFFFFFFFFFFFFFFFFFFFF
\begin{figure*}
  \includegraphics[width=7.5cm]{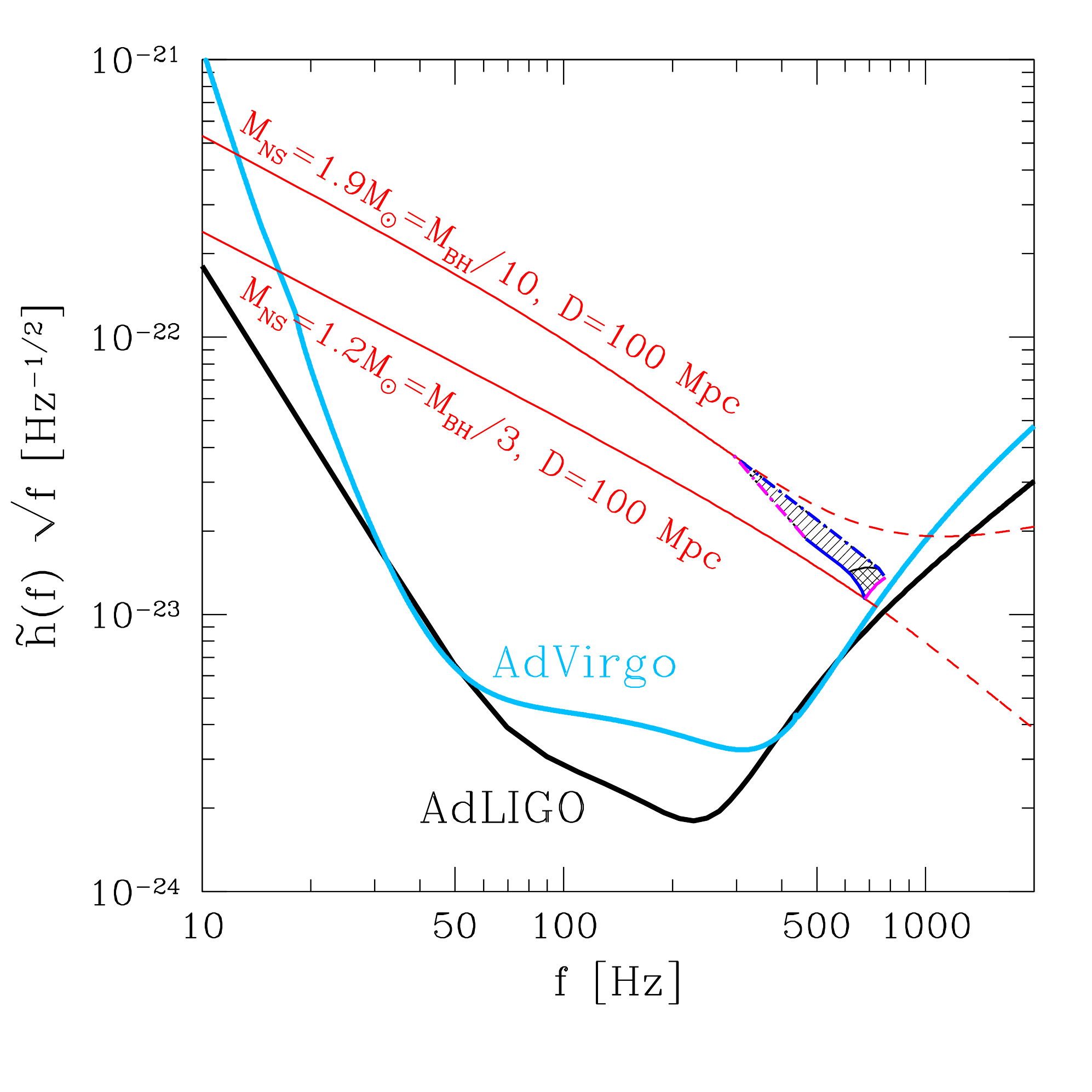}%figs/strain_vs_fe}%
  \hskip 1.0cm
  \includegraphics[width=7.5cm]{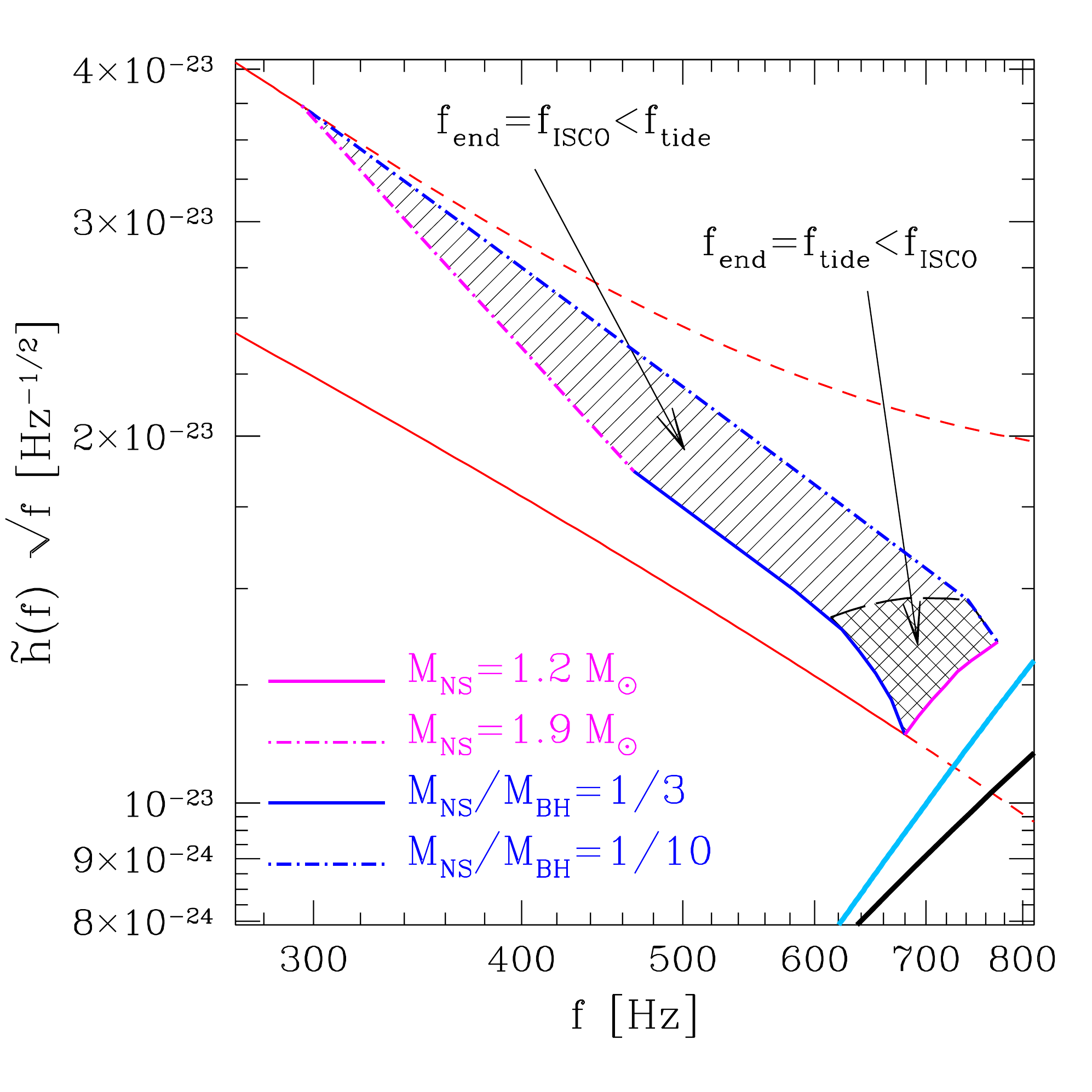}%figs/strain_vs_fe_zoom_legend}%
  \vskip -0.5cm
  \caption{Tracks of inspiralling BH-NS binaries at $100\,$Mpc, for
    $a=0$ and a PS EOS (red lines), and their position with respect to
    the sensitivities of AdLIGO (black line) and AdVirgo (light blue
    line). To avoid cluttering, we show only the strongest and the
    weakest signals, which refer to a $1.9\,M_\odot$ and a
    $1.2\,M_\odot$ NS, respectively, in binaries with mass ratio
    $q=0.1$ and $q=1/3$. Each track is terminated at
    $(f_{\text{end}},\tilde{h}(f_{\text{end}})\sqrt{f_{\text{end}}})$,
    the red dashed continuations are shown only for reference. The
    shaded region is the one spanned by all the termination points
    obtained by varying $M_\text{NS}$ and $q$. A magnification reporting
    the lines at constant $M_{\text{NS}}$ and $q$ that limit the shaded
    area is shown in the right panel.
    \label{FIG:signals}}
  \vskip -0.5cm
\end{figure*}
%FFFFFFFFFFFFFFFFFFFFFFFFFFFFFFFFFFFFFFFFFFFFFFFFFFFFFFFFFFFFFFFFFFFFF

\noindent\emph{Methodology.~}Early inspiral tidal effects may be
effectively described in terms of a single EOS-dependent tidal
deformability parameter $\lambda$~\cite{Flanagan08, Damour:2009,
  Hinderer09, Damour:2009wj} (also indicated as
$\mu_2$~\cite{Damour:2009, Damour:2009wj}), defined as the ratio of
the quadrupole deformation induced on a star and the static external
quadrupolar tidal field inducing the deformation, in this case the
tidal field of its companion. The tidal deformability parameter is
calculated via a linear axisymmetric $\ell = 2$ static perturbation
around the centre-of-mass of the star, along the axis connecting the
two companions; it depends on the EOS via the NS radius, $R$, and the
$\ell=2$ apsidal constant, or Love number, $k_2$, through the relation
$\lambda=2k_2R^5/(3G)$~\cite{Mora04, Berti08, Hinderer08}. The
$\ell=2$ tidal deformation is expected to be the dominant source of
EOS-dependent modifications to the inspiral phase evolution up to the
last few orbits of the inspiral prior to the merger, when the
mass-shedding takes over; higher-order in $\ell$ and $\lambda$,
nonlinear, and viscous dissipation corrections are all considerably
smaller~\cite{Hinderer09}. The imprint of $\lambda$ on the inspiral of
binary NSs was investigated in~\cite{Hinderer09}, which concluded
that, at a distance of $100\,$Mpc and using the inspiral below
$450\,$Hz, AdLIGO would be only be able to constrain $\lambda$ for an
extremely stiff EOS.

In order to estimate the GW phase accumulated during the inspiral of
BH-NS binaries due to tidal effects, we follow~\cite{Flanagan08,
  Hinderer09}, with the obvious difference that only one NS provides
the tidal contribution. We also include the first post-Newtonian
($1$PN) order tidal corrections calculated
in~\cite{Vines2011}. Finally, while in~\cite{Hinderer09} the phase
evolution was truncated at $450\,$Hz (roughly $80\%$ of
$f_{\text{ISCO}}$ for two NSs with \hbox{$M_{\text{NS}}=1.4\,M_\odot$}
and $R=15\,$km), we here set the value of the truncation frequency
$f_\text{end}$ to be the smaller of $f_{\rm tide}$ and
$f_\text{ISCO}$. The latter is the frequency yielded by the minimum of
the PN energy function (see below), while $f_{\rm tide}$ is determined
using the relativistic toy model recently discussed
in~\cite{Pannarale2010}, which reproduces numerical-relativity results
for $f_{\rm tide}$ within $\lesssim 1$\%. In such a model the NS is
described as an ellipsoid which deforms during the inspiral under the
effect of the BH tidal field and its internal forces; this is
consistent with the approach of~\cite{Hinderer09}, since only the
$\ell=2$ tidal deformations are considered.

We contrast the point-particle description of the quadrupole GW phase
($\phi_\text{PP}$) with the one obtained for a deformable, finite-size
NS ($\phi_\lambda$) by integrating the tidal corrections to
$d\phi/dx$, where $x$ is the PN expansion parameter, up to 1PN
(relative) order~\cite{Vines2011}, \ie we compute $\Delta\phi \equiv
\phi_{\text{PP}} - \phi_{\lambda}$ (which is positive, due to the
attractive nature of the tidal coupling). We moreover determine the
quadrupole mode of the gravitational radiation in the frequency domain
$\tilde{h}(f)$. We use the inspiral part of spinning binary hybrid
waveforms [Sect. III of~\cite{Santamaria2010}], augmenting the
TaylorF2 phase $\psi$ with $\lambda$-dependent tidal corrections up to
1PN (relative) order [Eq.~(3.9) in~\cite{Vines2011}]. The PN
expressions in~\cite{Santamaria2010} are derived from a $3$PN
expansion of the binding energy of the system and a $3.5$PN (relative
order) expansion of the GW energy flux [Eqs.~(3.3)-(3.4)
in~\cite{Santamaria2010}], within the commonly used stationary phase
approximation~\cite{Sathyaprakash_B:91}; spin terms are included up to
$2.5$PN order in phase and $2$PN order in amplitude. We use the
TaylorT4~\cite{Damour00a} Description to replace the $dx/dt$ in the
expression of the amplitude $A$.  Even though leading-order tidal
contributions are of $5$PN order, as opposed to the point-particle
terms which are kept up to $3$PN or $3.5$PN order, their coefficients
``compensate'' the smallness due to their PN order, making them
comparable to $3$PN and $3.5$PN terms (see~\cite{Mora04} and
~\cite{Damour:2009wj} for detailed discussions). We perform all
integrations in the frequency domain, from $f_\text{start}=10\,$Hz
(roughly the low-frequency cutoff of AdLIGO/AdVirgo) to
$f_{\text{end}}=\min (f_\text{tide},f_\text{ISCO})$.

The free parameters of our model are the NS (barotropic) EOS, the NS
mass $M_\text{NS}$, which ranges between $1.2\,M_\odot$ and the
EOS-dependent maximum value $M_\text{max}$, the binary mass ratio
$q\equiv M_\text{NS}/M_\text{BH}$, which we vary between $1/10$ and
$1/3$, and the dimensionless BH spin parameter $a$, which is taken to
be (anti-)aligned with the orbital angular momentum (equatorial
inspirals). To bracket the possible impact of the EOS on BH-NS
inspirals and their GW radiation, we consider three EOSs dubbed APR,
GNH3, and PS, respectively. Our choice is motivated by the fact that
the APR EOS~\cite{Akmal1998a} is based on nuclear many-body
calculations, may be favoured by the observations~\cite{Steiner2010},
and yields NSs with low $\lambda$; the GNH3
EOS~\cite{Glendenning1985}, on the other hand, is based on mean-field
theory and yields intermediate $\lambda$; finally, the ``liquid''
version of the PS EOS~\cite{PandharipandeSmith:1975} is somewhat dated
but yields rather high $\lambda$. The set chosen, therefore, covers
the relevant range of stiffness (APR being the softest and PS the
stiffest) and can be considered representative of a much larger sample
of EOSs.  Note that all of these EOSs have a maximum mass $1.93
\lesssim M_\text{max}/ \,M_\odot\lesssim 2.66$.

Figure~\ref{FIG:signals} illustrates the role played by
$f_\text{end}$. We consider the PS EOS and non-spinning BHs and show
the tracks of inspiralling BH-NS binaries at $100\,$Mpc (red solid
lines), along with the sensitivity curves of AdLIGO (black line) and
AdVirgo (light-blue line). The signal amplitudes are averaged over sky
location and relative inclination of the binary. We show explicitly
only the strongest and weakest signal, which refer to
$M_\text{NS}=1.9\,M_\odot$, $q=0.1$ and to $M_\text{NS}=1.2\,M_\odot$,
$q=1/3$, respectively. The tracks terminate at
$(f_\text{end},\tilde{h}(f_\text{end})\sqrt{f_\text{end}})$ and their
continuations as red dashed lines serve only as a reference. The
shaded region, which is magnified in the right panel, is the one
spanned by the termination point for all combinations of
$M_\text{NS}/M_\odot\in [1.2,1.9]$ and $q\in [0.1,1/3]$.

\begin{figure}%[]
  \includegraphics[height=8cm,angle=-90]{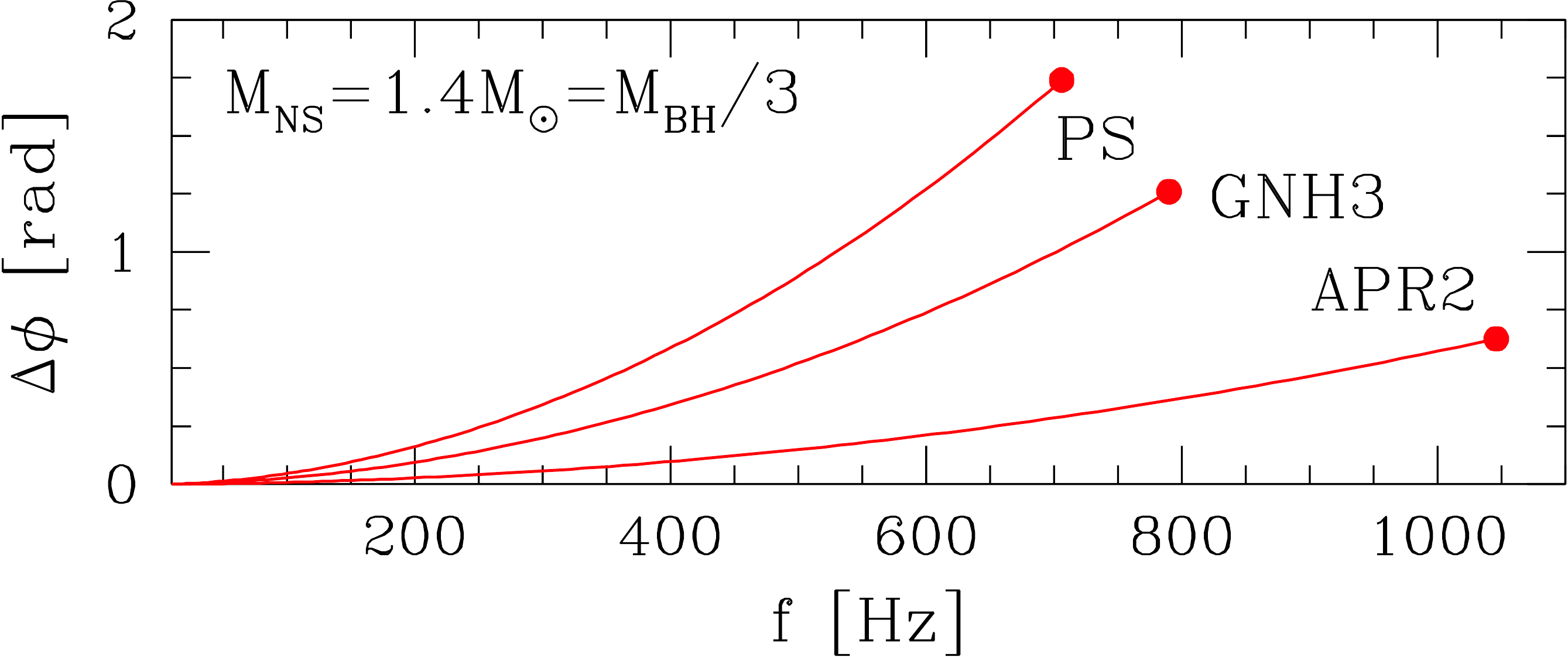}%figs/Deltaphi_vs_f}%
  \caption{Tidal distortion contribution to the quadrupole GW phase
    $\Delta\phi$ for the three representative EOSs.  The tracks end at
    $f_{\text{end}}$ and yield larger dephasings for stiffer EOSs,
    such as PS. \label{FIG:dephasing}}
  \vskip -0.5cm
\end{figure}

%%%%%%%%%%%%%%%%%%%%%%%%%%%%%%%%%%%%%%%%%%%%%%%%%%%%%%%%%%%%%%%%%%%%%%%
\noindent\emph{Dephasing and Overlaps.~}Once a binary
with parameters ($q,a,M_\text{NS},\lambda$) is selected, we compute
$\Delta\phi(f_\text{end})$. Overall, we find that $\Delta\phi$:
\textit{(i)~} is greater for bigger $\lambda$'s, \ie for more
deformable NSs (see Fig.~\ref{FIG:dephasing}); \textit{(ii)~}grows
with $q$, \ie for comparable masses; \textit{(iii)~} decreases as
$M_\text{NS}$ [\cf Eq.~(21) in~\cite{Hinderer09}]; \textit{(iv)~}
depends only weakly on the BH spin, since the only spin dependence may
come through $f_\text{end}$, but binaries with
$f_\text{end}=f_\text{tide}<f_\text{ISCO}$ are hardly affected, since
$f_\text{tide}$ is not very sensitive to $a$, while binaries with
$f_\text{end}=f_\text{ISCO}<f_\text{tide}$ are those with high
$M_\text{NS}$ and low $\lambda$, so that the gain or loss in
$f_\text{ISCO}$ does not modify $\Delta\phi$ significantly.

%FFFFFFFFFFFFFFFFFFFFFFFFFFFFFFFFFFFFFFFFFFFFFFFFFFFFFFFFFFFFFFFFFFFFF
%FFFFFFFFFFFFFFFFFFFFFFFFFFFFFFFFFFFFFFFFFFFFFFFFFFFFFFFFFFFFFFFFFFFFF

To determine whether the dephasings found may affect the detection of
BH-NS inspiral events, we compute, for each binary, the \emph{overlap}
between the point-particle model of the GW inspiral signal ($h_{\rm
  PP}$) and the one which includes tidal deformability effects
($h_\lambda$); this is the normalized inner product of the two
signals, maximised over time and phase shifts, \ie
$\mathcal{O}[h_{_{\rm PP}}, h_{_{\rm \lambda}}] \equiv
\max_{\{t_0,\phi_0\}}\frac{\langle h_{_{\rm PP}} | h_{_{\rm \lambda}}
  \rangle}{\sqrt{\langle h_{_{\rm PP}} | h_{_{\rm PP}} \rangle \langle
    h_{_{\rm \lambda}} | h_{_{\rm \lambda}} \rangle}}\,$,
where the inner product is
$ \langle h_{_{\rm PP}} | h_{_{\rm \lambda}} \rangle \equiv 4 \Re
\int_{f_\text{start}}^{f_{\text{end}}} df \frac{\tilde{h}_{_{\rm
      PP}}(f) \tilde{h}_{_{\rm \lambda}}^*(f)}{S_h(f)}\,$,
$S_h(f)$ being the noise power spectral density of a chosen detector.
Note that we are implicitly assuming that the waveforms including
tidal effects are the ``real'' signals and treating the point-particle
waveforms as the templates used to detect them.

\begin{figure}%[]
  \includegraphics[width=8cm]{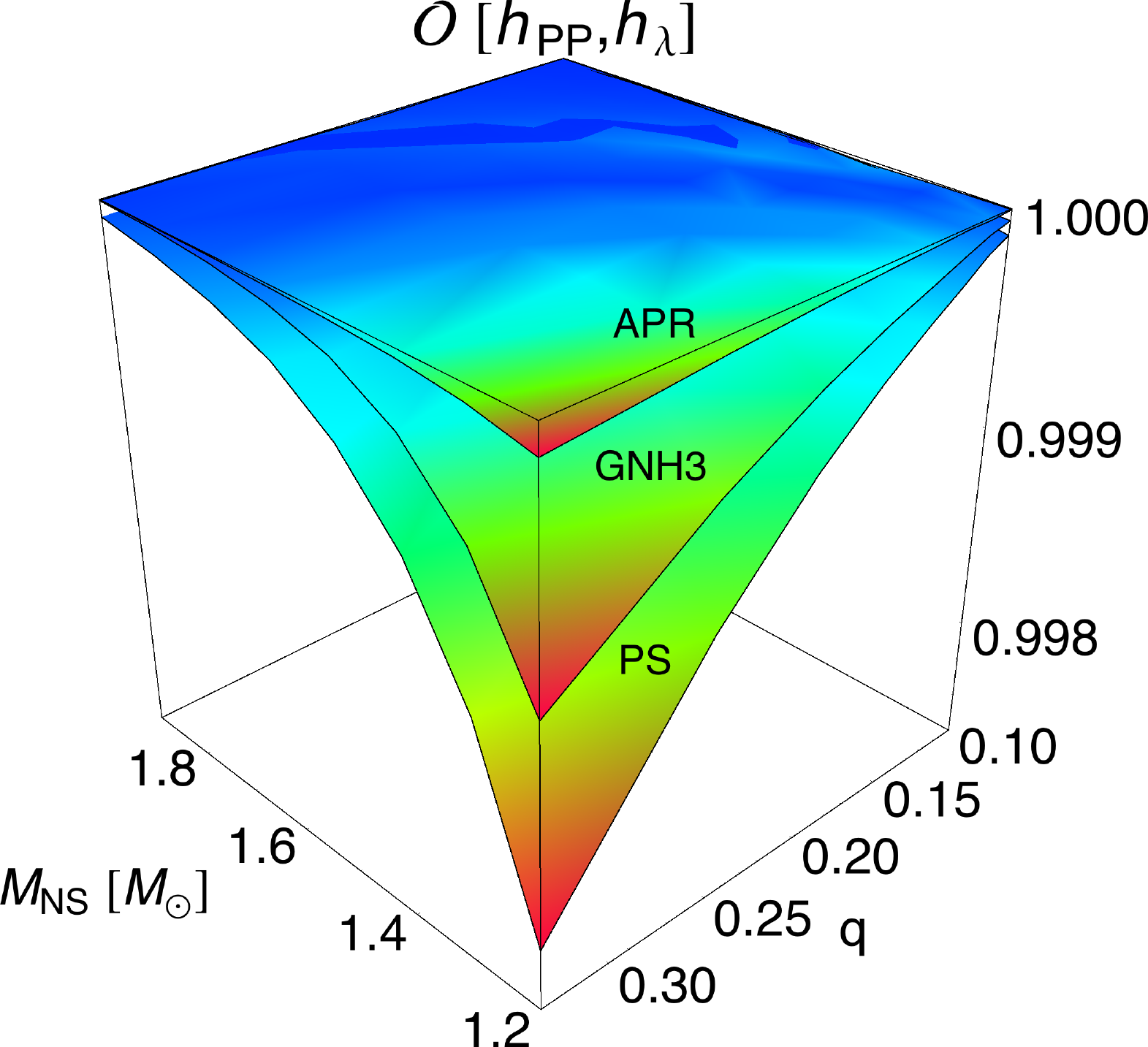}%figs/All3Overlaps}%
  \caption{Overlaps between PN waveforms for BH-NS binary systems
    modelled as point-particles (``PP'') and with the inclusion of
    tidal distortion effects (``$\lambda$''). The overlap is
    calculated for the AdLIGO detector and for non-spinning BHs. 
    \label{FIG:overlaps}}
  \vskip -0.5cm
\end{figure}

Our results for the three EOSs considered and a BH with $a=0$ are
shown in Fig.~\ref{FIG:overlaps} for AdLIGO. Note that for any EOS
choice and for any combination of the BH and NS masses, the overlap is
always greater than $0.997$, which corresponds to a $1$\% loss of
signals; this is true even for spins up to $a=1$. The smallest overlap
is given by the PS EOS combined with $M_{\text{NS}}=1.2\,M_\odot$,
$q=1/3$, and $a=1$ (inclusion of spin changes overlaps by $<
10^{-3}$). Hence, even if all binaries were to have these extreme
properties, the loss of signals would be less than $1$\%. All in all,
BH-BH inspiral templates will allow second-generation interferometers
to detect inspiralling BH-NS binaries with less than a $1$\% loss of
signals. Similar results hold for the third-generation detector
Einstein Telescope (ET)~\cite{Punturo:2010}, with a minimum possible
overlap of about $0.995$.

%FFFFFFFFFFFFFFFFFFFFFFFFFFFFFFFFFFFFFFFFFFFFFFFFFFFFFFFFFFFFFFFFFFFFF
%FFFFFFFFFFFFFFFFFFFFFFFFFFFFFFFFFFFFFFFFFFFFFFFFFFFFFFFFFFFFFFFFFFFFF

\noindent\emph{Measurability.~}Determining that the fraction of lost
signals is below $1$\% does not address the question of whether the
detected signals may be used to learn about the EOS. To address this
point, we consider a nominal detector-binary distance of $100\,$Mpc
and calculate the distinguishability as $\delta
h_{\text{PP},\lambda}\equiv \langle
h_\text{PP}-h_\lambda|h_\text{PP}-h_\lambda\rangle \gtrsim
2(1-\mathcal{O}[h_\text{PP},h_\lambda])\rho^2$, where $\rho^2=\langle
h_\text{PP}|h_\text{PP}\rangle\simeq\langle
h_\lambda|h_\lambda\rangle$ is the signal-to-noise-ratio (SNR) and we
neglected the term $\langle h_\text{PP}|h_\text{PP}\rangle - \langle
h_\lambda|h_\lambda\rangle \sim \! 10^{-4}$. Since we treat
$h_\text{PP}$ as the ``template'' and $h_\lambda$ as the ``signal'', a
necessary (but not sufficient) condition to distinguish between the
two is that $\delta
h_{\text{PP},\lambda}>1$~\cite{Lindblom:2008cm}. Clearly, the greater
$\delta h_{\text{PP},\lambda}$, the higher the chances of measuring
the tidal effects.

%FFFFFFFFFFFFFFFFFFFFFFFFFFFFFFFFFFFFFFFFFFFFFFFFFFFFFFFFFFFFFFFFFFFFF
\begin{figure}%[]
  \includegraphics[width=7.5cm]{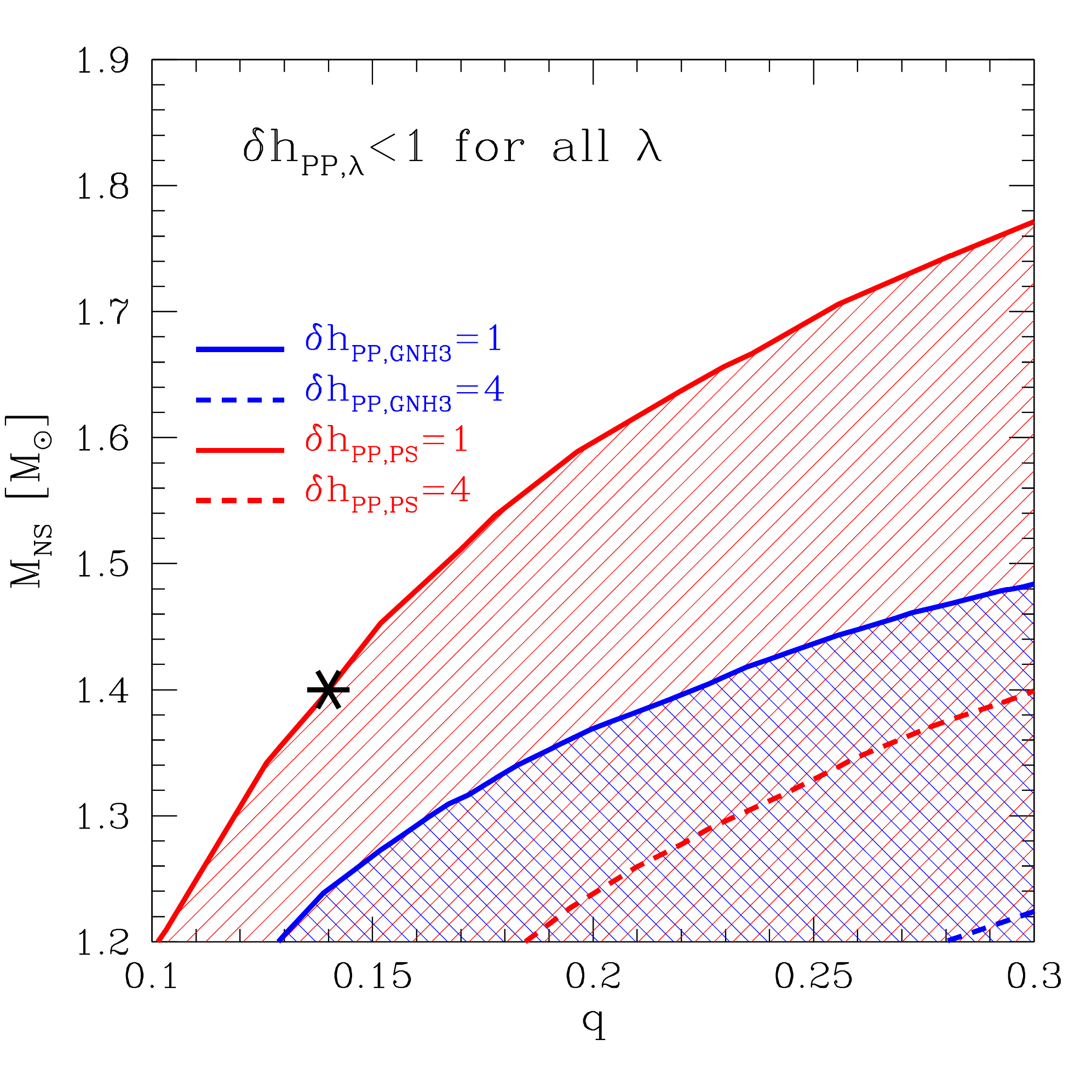}%figs/delta_h}%
  \vskip -0.5cm
  \caption{Distinguishability $\delta h_{\text{PP},\lambda}$ by AdLIGO
    for binaries at $100\,$Mpc and with $a=0$. In the white area
    $\delta h_{\text{PP},\lambda}<1$ for all EOSs, in the red-shaded
    one $\delta h_{\text{PP,PS}}> 1$, and in the blue-shaded one
    $\delta h_{\text{PP,GNH3}}> 1$. Contours for
    $h_{\text{PP},\lambda}=1, 4$ are shown for both EOSs. The black
    star marks a canonical $10\,M_\odot$-$1.4\,M_\odot$ BH-NS
    binary. \label{FIG:distinguishability}}
  \vskip -0.5cm
\end{figure}
%FFFFFFFFFFFFFFFFFFFFFFFFFFFFFFFFFFFFFFFFFFFFFFFFFFFFFFFFFFFFFFFFFFFFF

In Fig.~\ref{FIG:distinguishability}, we consider AdLIGO and report
$\delta h_{\text{PP},\lambda}$ for binaries with $a=0$. In calculating
the distinguishability, we average the signal amplitudes both over sky
location and over relative inclination of the binary. Note the
existence of a region where $\delta h_{\text{PP},\lambda}<1$ for all
EOSs (white area) in the space of parameters ($q, M_\text{NS}$); this
indicates that the inspiral of BH-NS binaries falling in this region
will not be distinguishable from a BH-BH inspiral. For larger mass
ratios and smaller NS masses, $\delta h_{\text{PP},\lambda}$
increases, becoming equal to $1$ first for the PS EOS (red-shaded
area) and then for the less stiff GNH3 EOS (blue-shaded area). The
maximum value of $\delta h_{\text{PP},\lambda}$ is $\sim \! 10$ ($\sim
\! 5$) for the PS (GNH3) EOS. Note that the black star pin-points a
canonical $10\,M_\odot$-$1.4\,M_\odot$ BH-NS binary and that, for the
popular APR EOS, $\delta h_{\text{PP,APR}}< 1$ for any ($q,
M_\text{NS}$). However, an APR(-like) EOS may become measurable if the
binary is optimally oriented and/or is less than $100\,$Mpc away. The
PS, GNH3, and APR EOS are (marginally) distinguishable by AdLIGO
within a range of $310$, $225$, and $75\,$Mpc, respectively, for a
$3.6\,M_\odot$-$1.2\,M_\odot$ BH-NS binary.

To test the robustness of our results, we repeated all calculations
after artificially decreasing $f_{\text{end}}$ by $20$\%. Even though
the overlaps (distinguishabilities) increased (decreased) a little
(\eg the minimum overlap increased by less than $10^{-3}$ and the
regions in which $h_{\text{PP},\lambda} \gtrsim 1$ were practically
unmodified), our conclusions remained unchanged, thus suggesting that
our results are robust even for non-negligible changes of
$f_{\text{end}}$.

\noindent\emph{Conclusions.~}The use of point-particle templates to
detect BH-NS inspirals leads to a loss of detected signals which is
below $1\%$ for both AdLIGO/AdVirgo and ET. Moreover, for binaries at
$100\,$Mpc, AdLIGO/AdVirgo will essentially be ``blind'' to tidal
effects if superdense matter follows an APR-like EOS and may be able to
reveal them only for NSs with a particularly stiff EOS and in
large-mass-ratio binaries. This scenario improves for a more sensitive
detector such as ET. In this case, the larger SNRs lead to a two-order
of magnitude gain in $\delta h_{\text{PP},\lambda}$, so that even soft
EOSs, like APR, are distinguishable.

A final remark should be made. The results presented here make use of
the most accurate PN expressions available to date and the comparisons
made with and without tidal corrections remove in part the problem of
systematic biases in the PN formulation, which could be large for
$q\lesssim 1$. Yet, they do not account for higher-order deformation
corrections which could amplify tidal
parameters~\cite{Damour:2009wj,Baiotti:2010}, and nonlinear responses
to the tidal field, such as those produced by crust fracturing or
resonant tidal excitation of stellar
modes~\cite{Kokkotas1995}. Although these contributions should yield
only fractional corrections to the already-small dephasings, either
increasing or decreasing them, it is only their proper inclusion that
will provide conclusive limits on the tidal effects that affect the
inspiral signal.

\noindent\emph{Acknowledgments.~} It is a pleasure to thank
E.~Berti, T.~Damour, T.~Hinderer, A.~Nagar, B.~S.~Sathyaprakash, and
A.~Tonita for carefully reading the manuscript. This work was
supported in part by the DFG grant SFB/Transregio~7, by ``CompStar'',
a Research Networking Programme of the European Science Foundation,
and by NSF grant PHY-0900735.

%%%%%%%%%%%%%%%%%%%%%%%%%%%%%%%%%%%%%%%%%%%%%%%%%%%%%%%%%%%%%%%%%%%%%%%%%%%%%% 
%%%%%%%%%%%%%%%%%%%%%%%%%%%%%%%%%%%%%%%%%%%%%%%%%%%%%%%%%%%%%%%%%%%%%%%%%%%%%% 
\bibliographystyle{apsrev4-1-noeprint}
\bibliography{aeireferences}
%%%%%%%%%%%%%%%%%%%%%%%%%%%%%%%%%%%%%%%%%%%%%%%%%%%%%%%%%%%%%%%%%%%%%%%%%%%%%%%
%%%%%%%%%%%%%%%%%%%%%%%%%%%%%%%%%%%%%%%%%%%%%%%%%%%%%%%%%%%%%%%%%%%%%%%%%%%%%%%
\end{document}